\documentclass[a4paper,11pt]{article}
\usepackage{pos}

\usepackage{graphicx}
\usepackage{amsmath,esint}
\usepackage{lineno}
\usepackage{subfig}
\usepackage{wrapfig}
\usepackage{setspace}
\newcommand{\spicecore}{{SPICEcore }}

\title{Design, performance, and analysis of a measurement of optical properties of antarctic ice below 400 nm}

\ShortTitle{UV calibration device}

\author{The IceCube Collaboration \\{\normalsize \normalfont(a complete list of authors can be found at the end of the proceedings)}}

\emailAdd{jannes.brostean-kaiser@icecube.wisc.edu}

\abstract{
The IceCube Neutrino Observatory, located at the geographic South Pole, is the world's largest neutrino telescope, instrumenting 1 km$^3$  of Antarctic ice with 5160 photosensors to detect Cherenkov light. For the IceCube Upgrade, to be deployed during the 2022-23 polar field season, and the enlarged detector IceCube-Gen2 several new optical sensor designs are under development. One of these optical sensors, the Wavelength-shifting Optical Module (WOM), uses wavelength-shifting and light-guiding techniques to measure Cherenkov photons in the UV range from 250\,nm to 380\,nm. In order to understand the potential gains from this new technology, a measurement of the scattering and absorption lengths of UV light was performed in the SPICEcore borehole at the South Pole during the winter seasons of 2018/2019 and 2019/2020. For this purpose, a calibration device with a UV light source and a detector using the wavelength shifting technology was developed. We present the design of the developed calibration device, its performance during the measurement campaigns, and the comparison of data to a Monte Carlo simulation.

\vspace{4mm}
{\bfseries Corresponding authors:}
Jannes Brostean-Kaiser$^{1*}$\\
{$^{1}$ \itshape DESY Zeuthen, D-15738 Zeuthen}\\[4mm]
$^*$ Presenter
\FullConference{37$^{\rm{th}}$ International Cosmic Ray Conference (ICRC 2021)\\
		July 12th -- 23rd, 2021\\
		Online -- Berlin, Germany}

}
\begin{document}
\maketitle
%\linenumbers

\section{Introduction / Wavelength-shifting Optical Module}\label{Wavelength Shifting Optical Module}  
The IceCube Neutrino Observatory is a cubic-kilometer detector installed in the ice at the geographic South Pole at depths between 1,450\,m and 2,450\,m \cite{detector:paper}.
The detector was completed in 2010. 
To reconstruct direction, energy, and flavor of interacting neutrinos the Cherenkov radiation, emitted by charged secondary particles, is measured. 
 
To improve the reconstruction of low energy neutrinos and the calibration of the instrumented ice, the IceCube Upgrade will be deployed in the austral summer 2022-2023. 
Seven additional strings will be deployed, including several types of novel optical modules. 

Several of the new modules under development, are designed to measure Cherekenkov radiation in the UV range. This improves the sensitivity of the modules since the number of emitted Cherenkov photons is proportional to one over the wavelength squared. 

One of these UV-sensitive detectors is the Wavelength-shifting Optical Module (WOM)\cite{WOM:2017}. 
%, which is currently in a prototype state  76 cmlong and approximately10.6 cmi
The WOM consists of a 76\,cm long transparent (PMMA or quartz glass) tube with 10.6\,cm diameter. The tube is coated with a wavelength-shifting paint \cite{WOM2021} and connected to two photomultiplier tubes (PMTs), one on each side. The paint absorbs photons with a wavelength between 250\,nm and 400\,nm and reemits them at roughly 420\,nm. The reemitted photons are guided via total internal reflection to one end of the tube and are detected by the PMTs. 

%In addition to the sensitive range, the WOM has due to smaller PMTs and higher effective area a much higher signal-to-noise ratio than the current optical modules. 

\section{Ice Properties}
To understand the potential improvement of new optical modules, the surrounding material has to be calibrated in the sensitive range. The Antarctic ice originates in compacted snow turning to ice over long times.
%The Antarctic ice originates in compacted snow which turns over a long time to ice.% Since this process is unique and not reproducible in a laboratory an in-situ measurement is necessary.
To measure scattering and absorption specifically, an in-situ measurement device, the UV calibration device (UV logger) has been built. 

\subsection{Absorption}
In the visible spectrum down to 300\,nm, the ice is mostly transparent, with absorption and scattering driven by impurities in the ice like dust, mineral, or soot \cite{RemoteSensingOfDust}. 
In the very deep UV range a strong absorption occurs, the ``Urbach tail'' \cite{Urba:1953}. The exact cutoff wavelength is yet unknown but believed to be below 200\,nm \cite{Mint:1971}.

\subsection{Scattering}
Using the AMANDA detector the scattering and absorption coefficient could be calibrated down to 337\,nm. Above 1300\,m depth the scattering is dominated by small air bubbles converting to craigite in the IceCube depth range due to the ice pressure \cite{Acke:2006}. 
Below this so-called bubble-dominated region, the photons scatter on aforementioned impurities. The particles have varying radii between a few nanometers and several micrometer \cite{RemoteSensingOfDust}, which results in a mixture of Rayleigh and Mie-Scattering. 

%\begin{figure}
%    \centering
%\includegraphics[width=0.35\textwidth]{ICAbs.png} \hfill 
%\includegraphics[width=0.6\textwidth]{ICScat.png}
%\captionof{figure}{a) Absorption and b) effective scattering coefficient of light in different wavelengths in the antarctic ice, measured by IceCube and AMANDA \cite{Acke:2006} \label{properties}}
%\end{figure} 

\section{In-situ measurement in the \spicecore hole}
The in-situ measurements were done in the South Pole ice core hole (\spicecore hole). It is an open borehole at about 1\,km distance from the IceCube array with a depth of 1750\,m \cite{spice} and 126 mm diameter. During the drilling process, the hole was filled with Estisol 140, a synthetic ester fluid that stays liquid in the South Pole environment. As its density is very similar to the surrounding ice, it prevents the hole from collapsing and keeps the hole open for calibration measurements. 

To measure in an open hole, the measurement device has to be the light emitter and detector at the same time. The light is sent out into the ice in nanosecond short pulses. The detector records the arrival time of the back-scattered photons. 
This time distribution can later be compared to simulation to obtain the ice properties. 
Early simulations suggest that the rising edge of the distribution is driven by the scattering coefficient, while the tail of the distribution is driven by the absorption coefficient. These effects are visible in the Figures \ref{TestCurves} a) and b). 
%From the time distribution, one can derive the properties of the surrounding ice.

Since a measurement with emitter and detector at the same place is more sensitive to backward scattering than forward scattering, an additional future task will be the comparison between this scattering measurement and former measurements with large detectors as IceCube or AMANDA. 

In addition to the UV Calibration device several other in-situ measurements took place in the two seasons as the Luminescence Logger \cite{Anna:2019}, the Camera System \cite{Cam2021} and the dust logger \cite{DustLogger}.

\section{Optimized UV calibration device}

%\begin{figure}
%    \centering
%\includegraphics[width=0.69\textwidth]{FullLog2.png} \hfill 
%\includegraphics[width=0.3\textwidth]{CrossSectionLabel.png}
% \begin{minipage}{0.5\textwidth}
% \centering
% a)
% \end{minipage}%
% \begin{minipage}{0.5\textwidth}
% \centering
% b)
% \end{minipage}
%\captionof{figure}{a) UV Calibration device with a detector, using PMTs, two open ones and four connected to wavelength shifting rods, a light source, capable of pulsing light with nanosecond pulse width and the read out electronic, stored in a quarz glass vessel with titanium endcaps and flanges. b) Cross section of the Calibration device. \label{Logger}.}
%\end{figure} 

\begin{figure}
\includegraphics[width=1\textwidth]{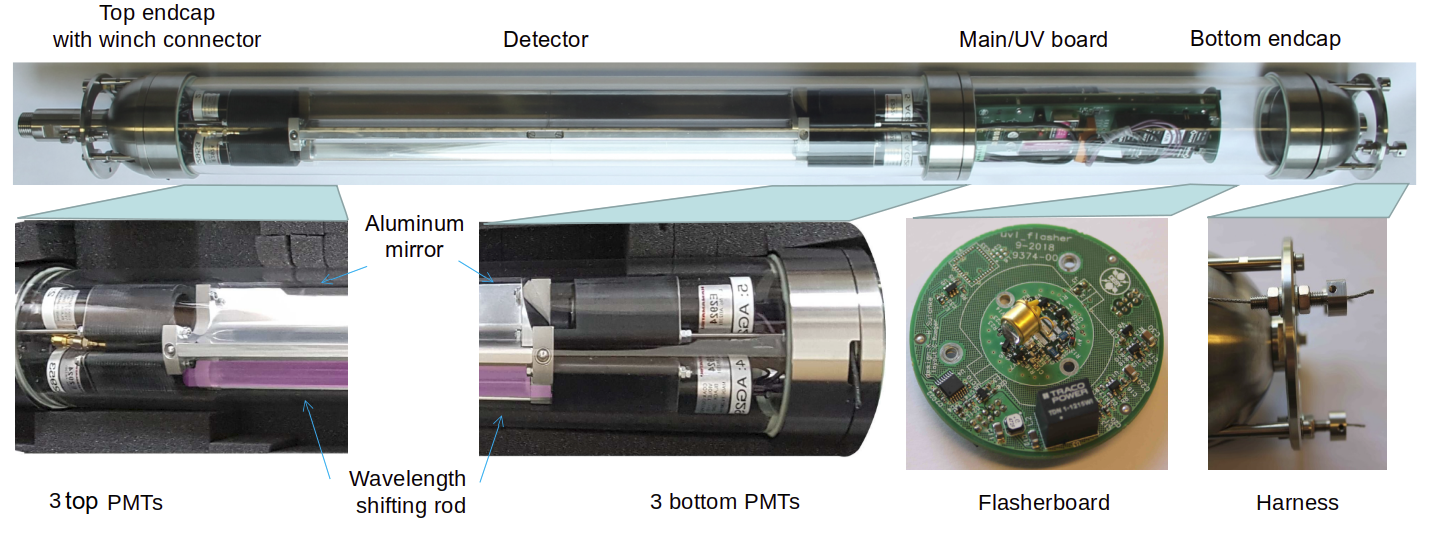}
\captionof{figure}{UV Calibration device with a detector, using PMTs, two open ones and four connected to wavelength shifting rods, a light source, capable of pulsing light with nanosecond pulse width and the read out electronic, stored in a quarz glass vessel with titanium endcaps and flanges \label{Logger}.}
\end{figure} 

The device, designed for this measurement consists of a LED-based light source with different wavelengths and a UV-sensitive detector. The detector is divided longitudinally into three segments by aluminum mirrors. Two PMTs are placed in every segment (six in total), one near the light source (bottom) and one on the top. In the two segments opposite of the LED, PMMA rods of 50\,cm length and 2\,cm diameter are connected to the PMTs. The rods are coated with a wavelength shifting paint, developed for the WOM. In the segment facing the same direction as the LED the PMTs are left open for direct photon detection. On the bottom PMT, an additional small mirror is placed to increase the sensitivity of photons with only a few scattering processes. Figure \ref{Logger} shows the full logger with all components. 

\newpage
\begin{wrapfigure}{r}[0cm]{7cm}
\centering
\includegraphics[width=0.35\textwidth]{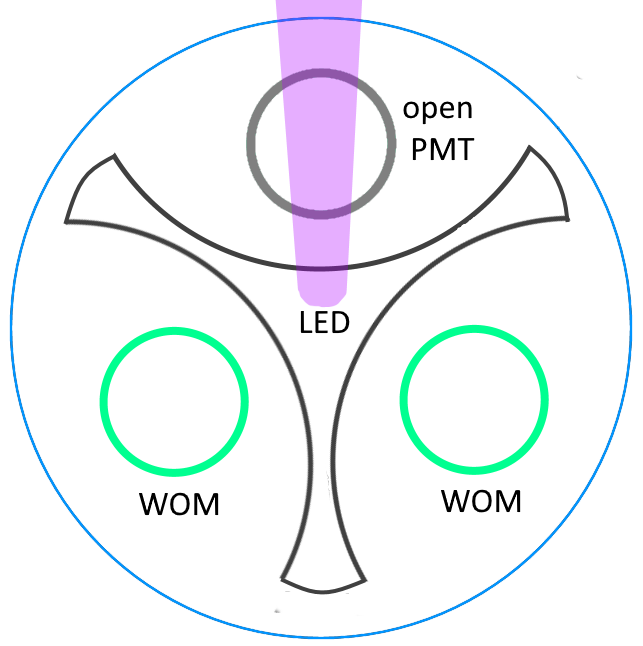} \captionof{figure}{Cross section of the UV calibration device with the WOMs and open PMTs sketched according to the LED emission angle \label{Cross section}}
\end{wrapfigure}

Most of the development and design have been done prior to the first measurement season and can be read up in previous works \cite{icrc2019}. Only the light source was altered between the two measurement seasons. 
The light source is based on flasher boards with one LED each.
In the two measurement seasons four flasher boards with wavelengths of 245\,nm, 278\,nm, 310\,nm and 370\,nm were used. The nanosecond light pulses are obtained using a Kapustinsky Pulser with adjustible light intensity. In the first measurement season an integrating sphere \cite{Pocam2021} was used to create a well-defined emission profile. For the second measurement season the integrating sphere was removed to increase the number of emitted photons. 
%\subsection{Detector}
%The detector was designed to detect photons in the range of 250\,nm to 370\,nm. 
%The detector is based on six PMTs\footnote{Hamamatsu R1924A (\href{https://www.hamamatsu.com/us/en/product/type/R1924A/index.html}{website}) operated at a gain of about $10^6$}, three on each side of the detector. The detector is divided longitudinal by aluminum mirrors into three segments. Two segments are holding a 50 cm long PMMA rod coated in wavelength shifting paint \cite{Hebe:2014}, connected to a PMT on each end of the rod. If a photon hits the wavelength shifting rod it gets absorbed, reemitted and is led via total reflection to the PMTs to be detected. 

%To find a solution since there is no wavelength shifting effect to give an additional time smearing effect on the data.  

% \subsection{Readout}
%The readout system is an FPGA-based data acquisition. It measures time differences between the trigger of the light source and the arrival of backscattered photons in each channel.
%Only one timestamp can be recorded for each PMT, per light pulse. Therefor the light intensity needs to be adjusted that on average only one photon gets scattered back into the detection unit. Photons arriving after the first photon would alter the waveform affecting the measurement of the arrival time of the first photon.
%As a cross check, for every thousand timestamps, one 1\,$\upmu$s long waveform is recorded with a Domino Ring Sampler.

\section{Measurements} 

\begin{wrapfigure}{l}[0cm]{7.8cm}
\centering
\includegraphics[width=0.5\textwidth
]{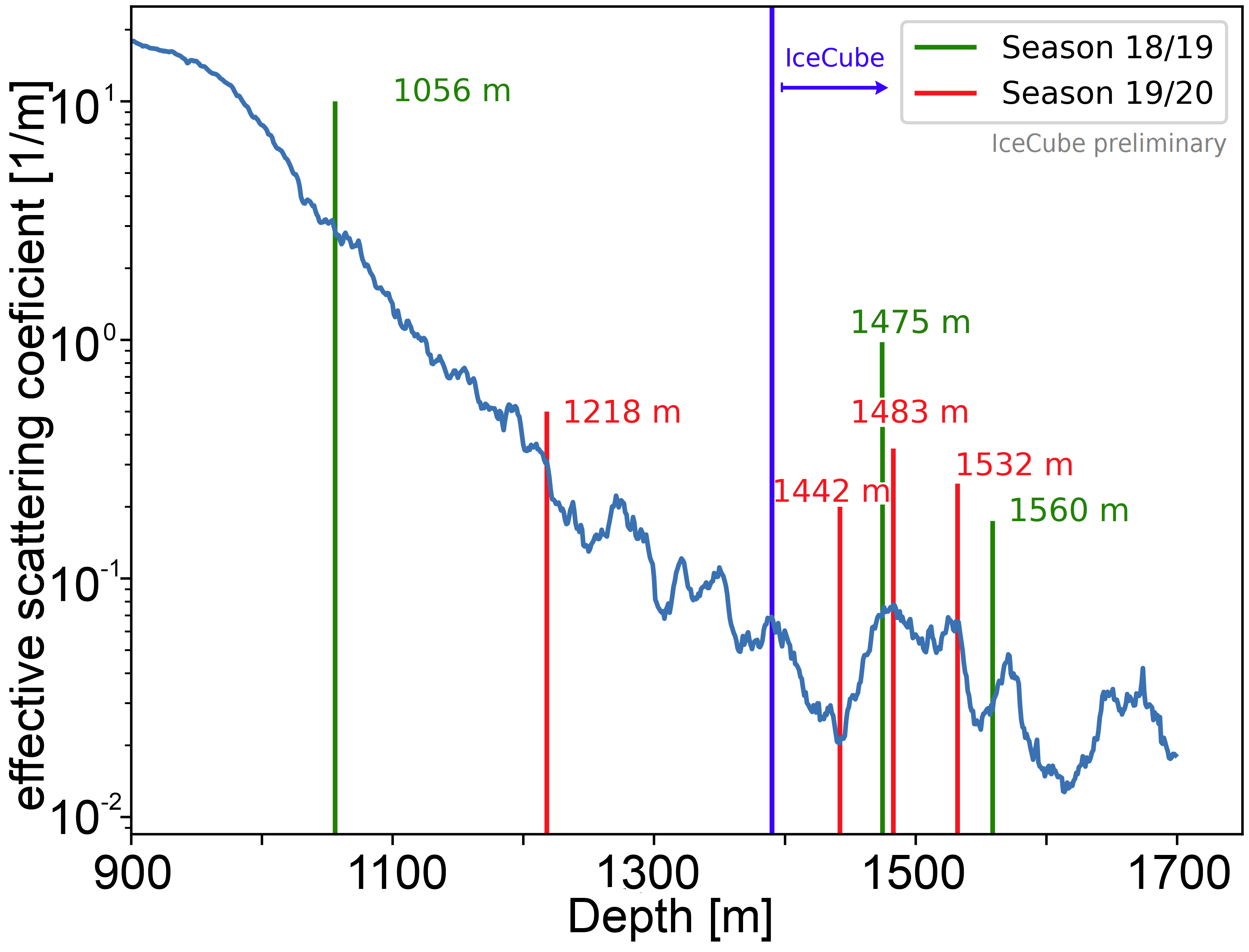} \captionof{figure}{All measurement depths of the two seasons, together with the effective scattering coefficients\cite{Acke:2006}, shifted to compensate the ice tilt between IceCube and SPICEcore. Depending on the depth the error of the ice tilt can increase up to 30\,m. \label{depth}}
\end{wrapfigure}

The measurements were done in two seasons with a total of 4 wavelengths at 7 depths in the ice. Figure \ref{depth} shows the measurement depths together with the expected scattering coefficients. 

\subsection{First measurement season}
In the austral summer 2018/2019 the first data set was collected on two days, at depths of 1056\,m, 1475\,m,  and 1560\,m, using both the 278 nm and 400 nm LED at each depth.  
Due to light intensity problems only the 278 nm LED provided useful data. During the whole measurement season one of the PMT channels, connected to a wavelength shifting rod did not record data. For some measurements the open PMTs picked up electric noise from the light source, but in every measurement at least 3 Channels recorded useful data. 

\subsection{Second measurement season}

The second measurement was performed in the austral summer 2019/2020. In total 4 measurement days were taken with three different flasher boards, where the flasher board with 250\,nm was used on two measurement days.
The measurements were done at depths of 1218\,m, 1442\,m, 1483\,m and 1532\,m.

The measurements with  250\,nm, 310\,nm and 370\,nm all provided useful data. For the 250\,nm measurement one channel connected to a wavelength shifting rod was not working. 

\subsection{Data preparation}
%Due to firmware problems in the saving process the resolution of the timestamps is reduced to 8\,ns and the there are some artifacts in the measurements. The timestamps are saved with 8\,ns resolution on a hard drive. 
To prepare the data for analysis it is represented in the form of histograms with 8\,ns bins (limited by a firmware bug), and cut to a time window from 80\,ns to 1050\,ns. The PMTs connected to the wavelength shifting rods are summed for each side, to have only two WOM channels, one for the bottom PMTs (the side nearer to the light source) next to the PMT with the mirror and one for the top PMTs (further away from the light source).

\begin{figure}
    \centering
\includegraphics[width=0.49\textwidth]{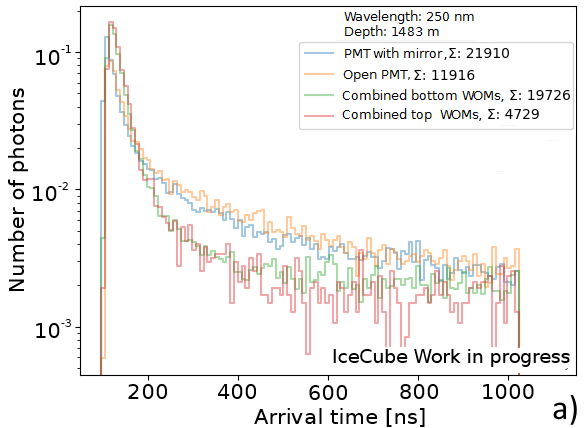} \hfill 
\includegraphics[width=0.49\textwidth]{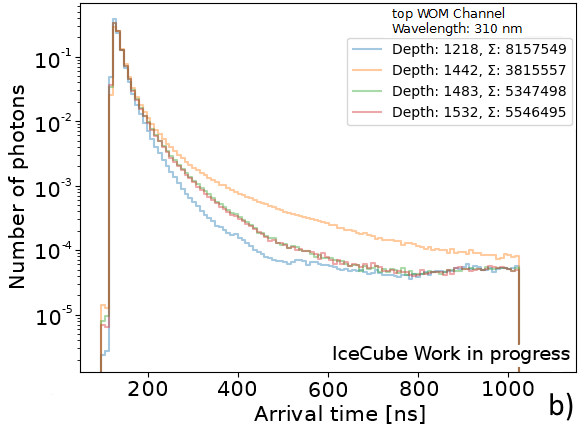}
% \begin{minipage}{0.5\textwidth}
% \centering
% a)
% \end{minipage}%
% \begin{minipage}{0.5\textwidth}
% \centering
% b)
% \end{minipage}
\captionof{figure}{Prepared example data sets of the measurements a) with 250\,nm at 1483\,m depth for all channels and b) with 310\,nm and the top WOM channel for all measured depths. \label{data}}
\end{figure}

Figure \ref{data} shows two sorted and prepared example datasets. (a) displays all channels of one measurement with a wavelength of 245\,nm at a depth of 1483\,m, (b) displays the top WOM channel for all measured depths with 310\,nm.
From these examples, it is evident, that the WOM channels have a larger time spread due to the wavelength shifting. Also the different depths have visible differences in the histograms.%, with the highest scattering at 1218\,m as expected.

%Since the emission profile and light intensity varies between the used LEDs, it is not practical to compare the different wavelengths at one depth directly to each other. 

\section{Data Analysis}
The analysis is done by comparing the experimental data to  Monte Carlo (MC) simulation with different absorption and scattering coefficients. The comparison to data is done using a binned maximum likelihood fit.

\subsection{Simulation}
The simulation models the experimental design in as much detail as possible. For the light emission, angular distribution, and wavelength spectrum of the LEDs datasheet values are interpolated. \\
The simulation follows the light path out of the calibration device through the quartz glass and Estisol into the ice using Fresnel equations. 
Every photon reaching the ice is assigned an absorption and scattering length sampled from random exponential distributions with the absorption and scattering coefficient as coefficients.
After each scattering length, a scattering angle is sampled and the photon receives a new direction and scattering length. After every scattering process, the traveled path length is integrated and compared to the absorption length. After passing the assigned absorption length in the ice the photon is counted as absorbed in the ice. The scattering angle %responsible for a new flight direction 
is highly dependent on the scattering model. For the simulation, Mie-Scattering was tested, but found to be impractical, since the experiment is mostly sensitive to backward scattering. Rayleigh scattering is used instead.
The angular distribution for Rayleigh scattering follows a $(1-cos\vartheta)^2$-distribution, with $\vartheta$ as the scattering angle. 

Photons scattered back to the detector again pass through the Estisol and quartz glass into the detector and are counted as detected when crossing a PMT or wavelength shifting rod. 
The transit time spread of the different detection ways was measured in the laboratory and is dependent on the position of the photon.

\subsection{Maximum Likelihood fit}
To analyze the measurements, the distribution of binned photon arrival times is compared to the simulation. 
The comparison is done by calculating a test statistic %over degrees of freedom 
$TS$ %dof
for every simulation according to the formula 
\begin{equation}
    TS = \sum_{i=1}^N \frac{(d_i - a_i\cdot N_d/N_a)^2}{d_i + a_i \cdot N_d^2/N_a^2}
\end{equation}
where $N$ is total number of bins in the measurement, $d_i$ and $a_i$ are the number of events in the bin $i$ for the measurement $d$ and the Monte-Carlo simulation $a$ and $N_d$ and $N_a$ are the total number of events in the measurement and Monte-Carlo simulation \cite{Barl:1993}. 
%If the number of photons in the simulation is much larger than in the dataset, the $TS/$dof would converge to 1 for the best fits. In this runs the data sets are by a factor 10-100 larger than the simulation sets which results in higher, but still stable $TS/$dof - values. 

\begin{figure}
    \centering
\includegraphics[width=0.49\textwidth]{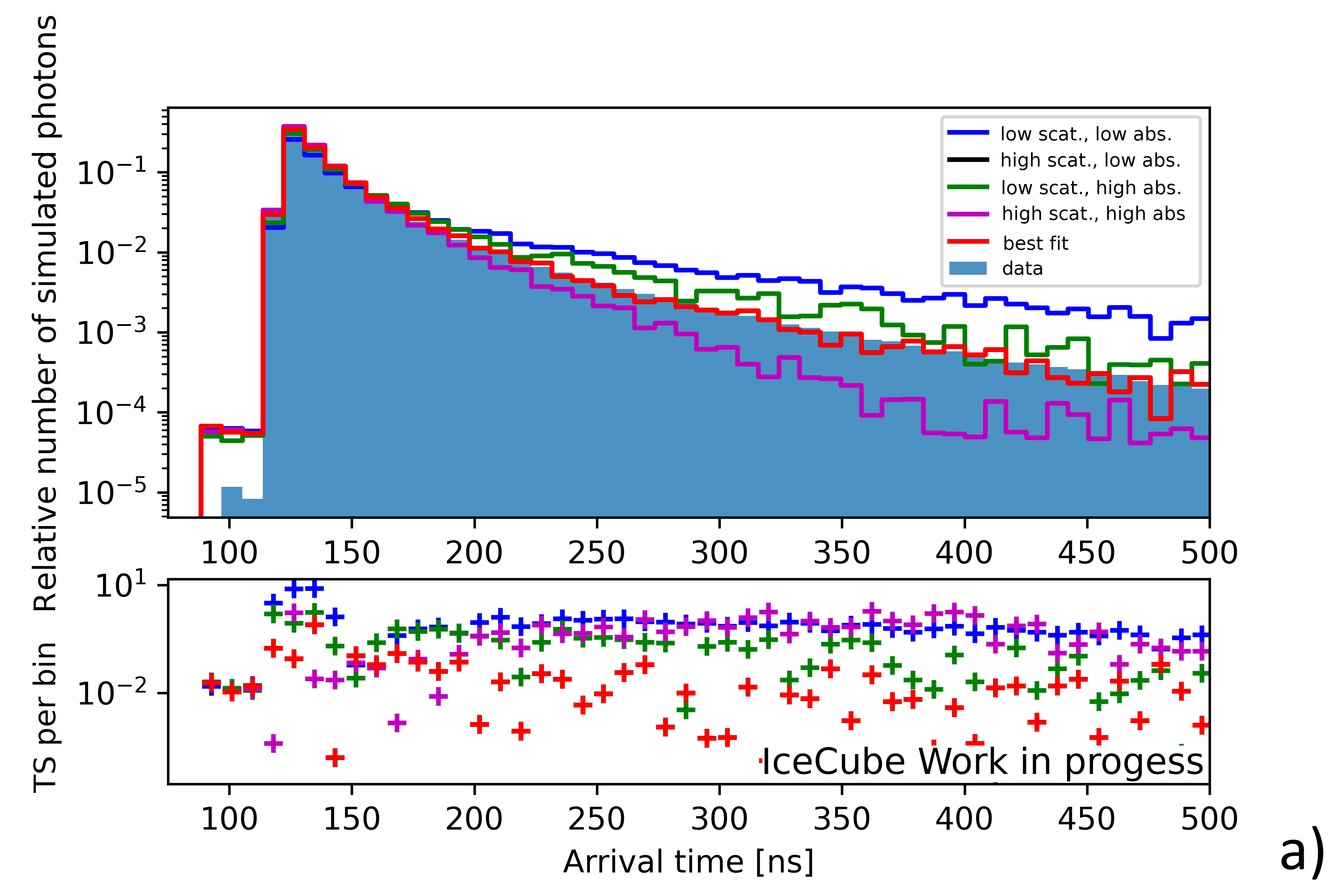} \hfill
\includegraphics[width=0.49\textwidth]{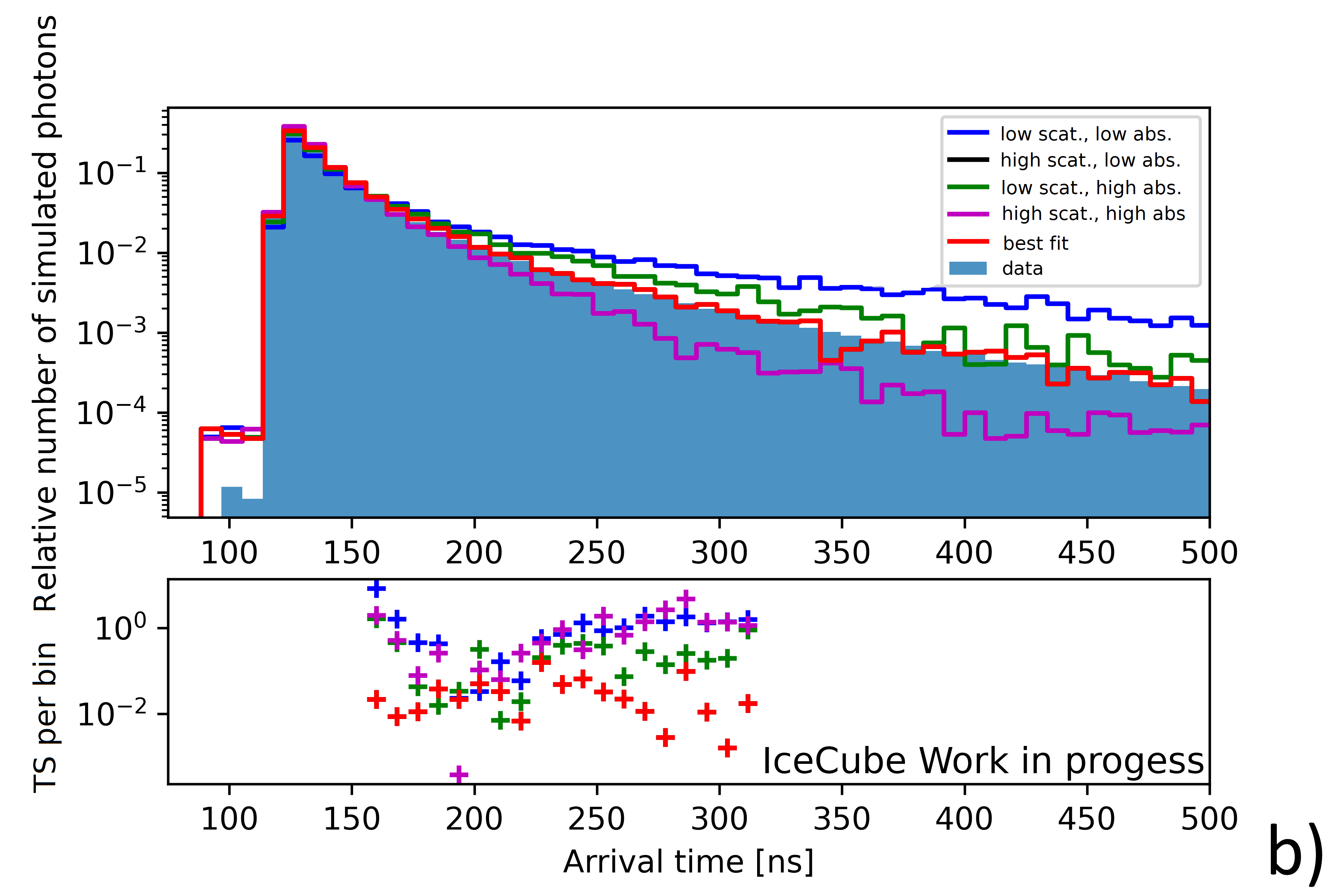}
% \begin{minipage}{0.5\textwidth}
% \centering
% a)
% \end{minipage}%
% \begin{minipage}{0.5\textwidth}
% \centering
% b)
% \end{minipage}
\captionof{figure}{ Dataset of a measurement with 5 simulations, 4 at the edges of the chosen parameter space and one best fit, a) for all bins with 10 or more entries, b) for a restricted time window of 150\,ns - 300\,ns. \label{TestCurves}}
\end{figure}

With this test statistic, a best fitting simulation with a given set of parameters can be found. Figure \ref{TestCurves} a) shows how the data of one depth, wavelength, and PMT-Channel connected to a WOM and five simulations are matching up. Four simulations are done with a set of high or low scattering and absorption parameters to show the boundaries of the chosen 2D scan. One simulation with a set of medium coefficients is shown in red and gives the best fit with the smallest calculated $TS$ %/$dof 
. Below the time distributions, the $TS$ per bin is plotted, so to understand the influence of each part of the distribution

To find a region of trustworthy minima the simulation with the lowest $TS$ %/$dof
is re-simulated and analysed 100 times to find a standard deviation $\sigma$. The true value for the parameters is supposed to lie inside an area where the difference of the $TS$ %/$dof
values to the minimum is smaller than $\sigma$, called the 1$\sigma$ region. 
This method is used to compensate for the limited simulation time. Since the number of simulated photons are smaller by a factor 10 to 100 it statistical error is mostly driven by the simulation instead of the measurement. This represents only the statistical error and not the systematic errors of the measurement. 

\subsection{Open issues}
\begin{figure}
    \centering
\includegraphics[width=0.49\textwidth]{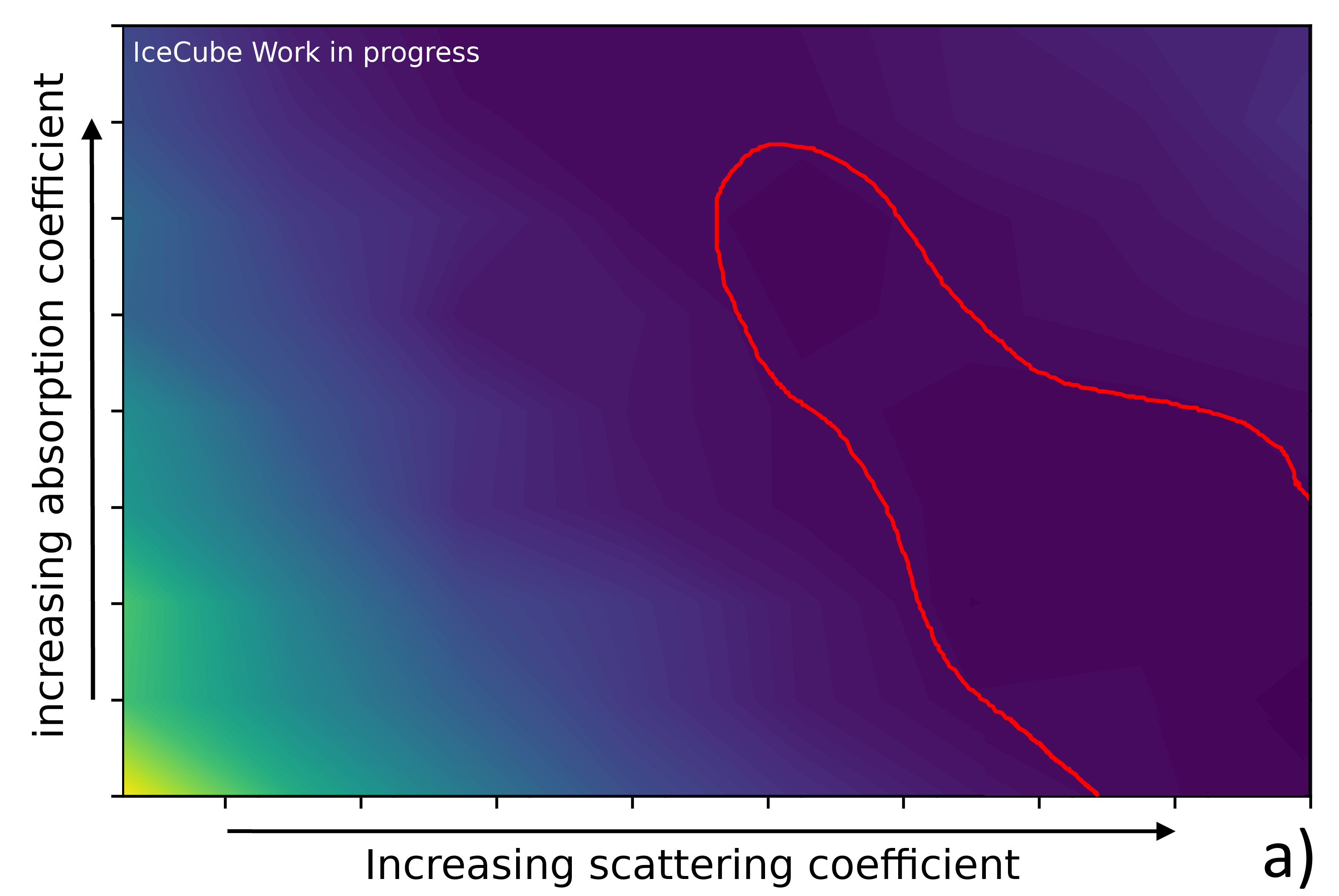}\hfill
\includegraphics[width=0.49\textwidth]{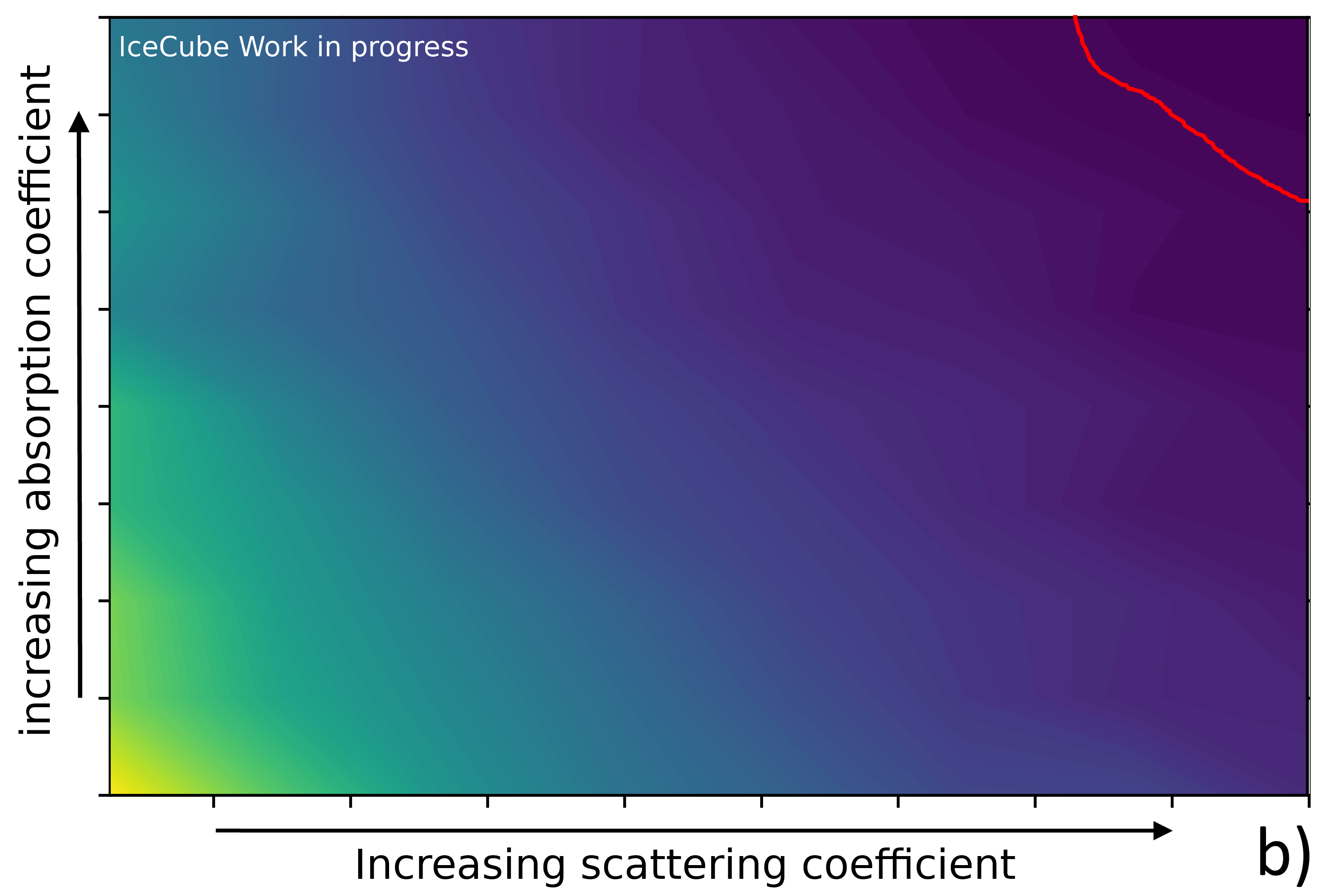}
% \begin{minipage}{0.5\textwidth}
% \centering
% a)
% \end{minipage}%
% \begin{minipage}{0.5\textwidth}
% \centering
% b)
% \end{minipage}
\captionof{figure}{ Simplified $TS$ %/$dof
grid of several simulated sets of parameters compared to one data set for two PMT-channels of the same measurement. \label{ContPlotAll}}
\end{figure}

The analysis returns a well defined minimum for each channel of the measurement, but there are still unsolved inconsistencies to be explained. Figure \ref{ContPlotAll} a) and b) show two simplified simulation grids of $TS$ %/$dof
calculations as a function of absorption and scattering.
%On the x-axis is an increasing simulated scattering coefficient and on the y-axis an increasing simulated absorption coefficient plotted. 
Both axes depict about 1 order of magnitude for each parameter. The red curve indicates the 1$\sigma$ region around the minimum. 

The first unexplained observation is the differences between the PMT-channels. Comparing the minima in Figure \ref{ContPlotAll} a) and b) the $\sigma$ regions are not overlapping. Therefore no definite minimum connecting all channels of one measurement has yet been found. 
This questions the correctness of the simulation and how well the experimental setup is understood.

Another concern is the size of the $\sigma$ region.
For some measurements as \ref{ContPlotAll} a) it covers almost the whole simulation grid. This and the form of the $\sigma$ region indicate a strong correlation of the two parameters. 
The choice of the scanned parameter space has to be therefore made very carefully to not have a minimum on the borders of the scanned area.
%This also means, that the calculated minimum can lie on the borders of the scanned area.

\begin{figure}
    \centering
\includegraphics[width=0.49\textwidth]{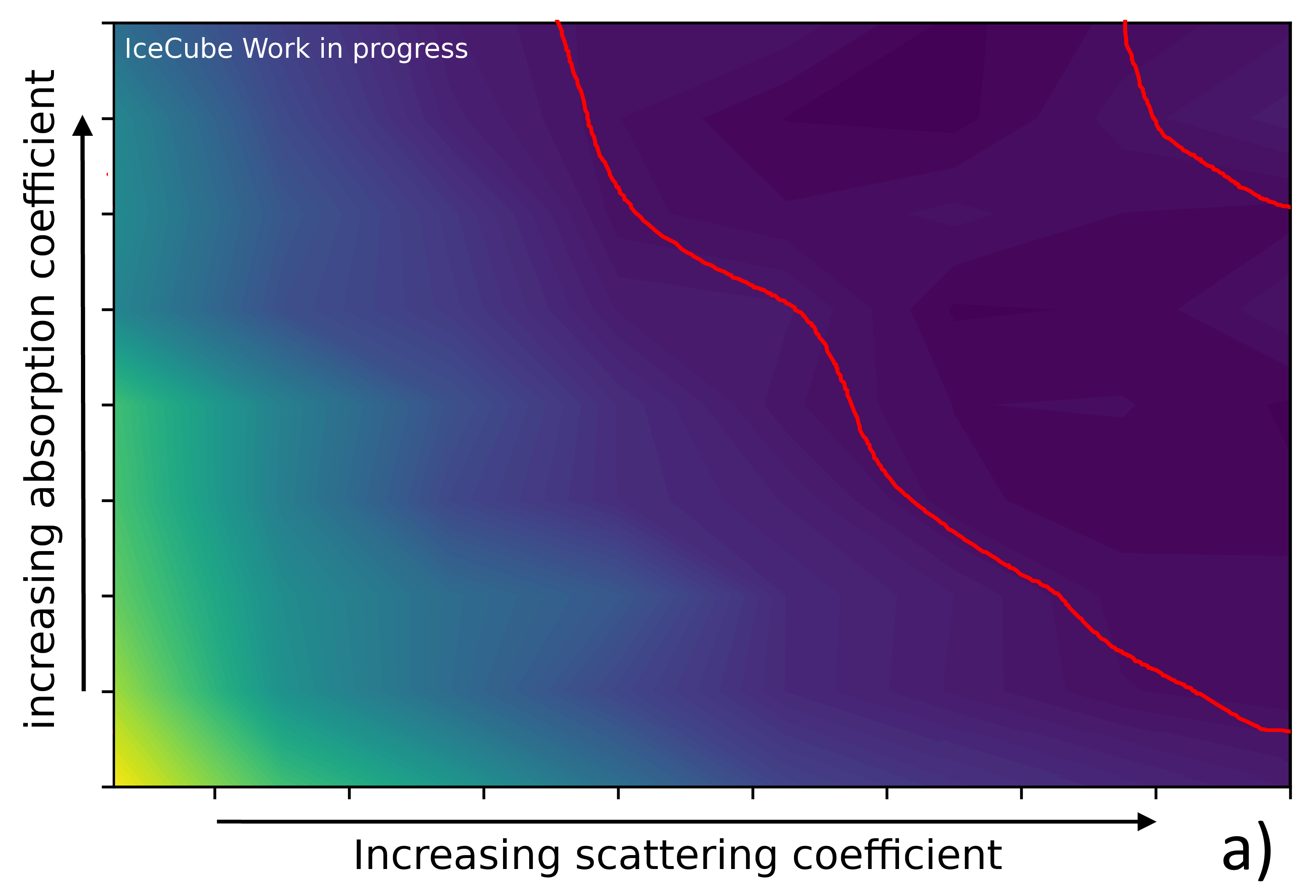} \hfill
\includegraphics[width=0.49\textwidth]{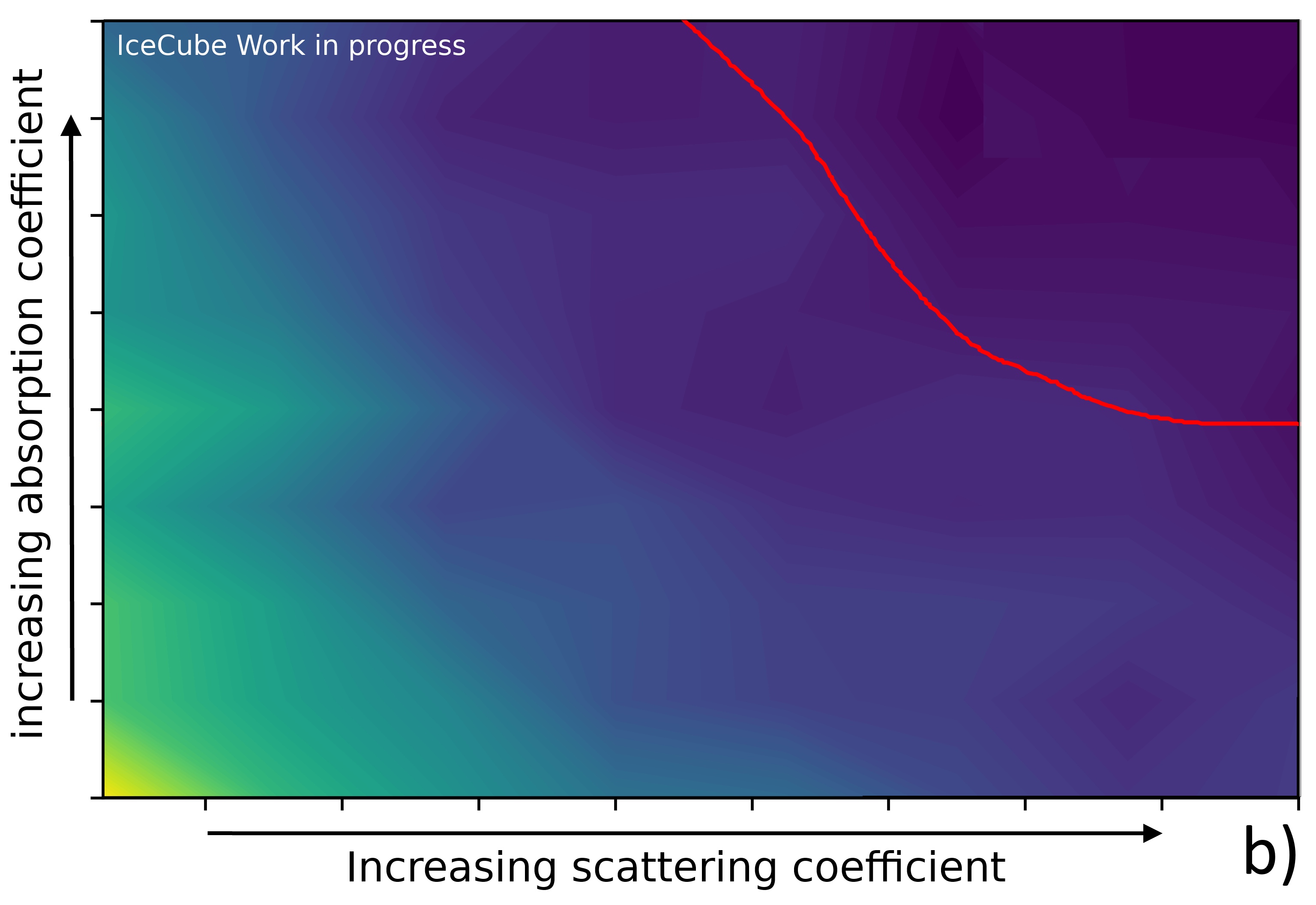}
% \begin{minipage}{0.5\textwidth}
% \centering
% a)
% \end{minipage}%
% \begin{minipage}{0.5\textwidth}
% \centering
% b)
% \end{minipage}
\captionof{figure}{ Simplified $TS$ %/$dof
grid of several simulated sets of parameters compared to one data set for two PMT-channels of the same measurement with a restricted time window. \label{ContPlotRest}}
\end{figure}

To decouple the two parameters the histograms are restricted to a time window of 150\,ns - 300\,ns, where the distributions are believed to be mostly absorption driven. Figure \ref{TestCurves} b)  again shows the best fit and several example simulations for this restricted time window. Figure \ref{ContPlotRest} gives again the simplified simulation grid with the $\sigma$ region around the minimum, showing still the same dependency of the two parameters. This leads to the conclusion that the two parameters are not easily decoupled and the final results could be a combined extinction parameter instead of independent absorption and scattering coefficients. 

\section{Outlook}
In the future, the focus will be on increasing the understanding of the experimental setup to understand and compensate for the differences in the measurement channels. This should lead to a combined minimum for each data set on each measured wavelength and depth, which can be compared to previous ice calibrations. 

\subsubsection*{Acknowledgements}
The author would like to thank the SPICEcore collaboration for providing the borehole, the US Ice Drilling Program, the Antarctic Support Contractor and the NSF National Science Foundation
for providing the equipment to perform the measurement and for their support at South Pole.

\bibliographystyle{ICRC}
\bibliography{ref}

\clearpage
\section*{Full Author List: IceCube Collaboration}

% \noindent \textbf{Note comment afterwards:} Collaborations have the possibility to provide an authors list in xml format which will be used while generating the DOI entries making the full authors list searchable in databases like Inspire HEP. For instructions please go to icrc2021.desy.de/proceedings or contact us under icrc2021proc@desy.de.\\

% \scriptsize
% \noindent
% first.author$^1$, 
% second.author$^2$, 
% third.author$^3$ % .... more names
% and 
% last.author$^{n}$ \\

% \noindent
% $^1$first.affiliation.
% $^2$second.affiliation. % .... more affiliation
% $^{m}$last.affiliation.

\scriptsize
\noindent
R. Abbasi$^{17}$,
M. Ackermann$^{59}$,
J. Adams$^{18}$,
J. A. Aguilar$^{12}$,
M. Ahlers$^{22}$,
M. Ahrens$^{50}$,
C. Alispach$^{28}$,
A. A. Alves Jr.$^{31}$,
N. M. Amin$^{42}$,
R. An$^{14}$
% ...
\\

\noindent
% ...
$^{12}$ Universit{\'e} Libre de Bruxelles, Science Faculty CP230, B-1050 Brussels, Belgium \\
% ...
$^{14}$ Department of Physics and Laboratory for Particle Physics and Cosmology, Harvard University, Cambridge, MA 02138, USA \\
% ...
$^{17}$ Department of Physics, Loyola University Chicago, Chicago, IL 60660, USA \\
$^{18}$ Dept. of Physics and Astronomy, University of Canterbury, Private Bag 4800, Christchurch, New Zealand \\
% ...
$^{22}$ Niels Bohr Institute, University of Copenhagen, DK-2100 Copenhagen, Denmark \\
% ...
$^{28}$ D{\'e}partement de physique nucl{\'e}aire et corpusculaire, Universit{\'e} de Gen{\`e}ve, CH-1211 Gen{\`e}ve, Switzerland \\
% ...
$^{31}$ Karlsruhe Institute of Technology, Institute for Astroparticle Physics, D-76021 Karlsruhe, Germany  \\
% ...
$^{42}$ Bartol Research Institute and Dept. of Physics and Astronomy, University of Delaware, Newark, DE 19716, USA \\
% ...
$^{50}$ Oskar Klein Centre and Dept. of Physics, Stockholm University, SE-10691 Stockholm, Sweden \\
% ...
$^{59}$ DESY, D-15738 Zeuthen, Germany \\
% ...

\subsection*{Acknowledgements}

\noindent
USA {\textendash} U.S. National Science Foundation-Office of Polar Programs,
U.S. National Science Foundation-Physics Division,
U.S. National Science Foundation-EPSCoR,
Wisconsin Alumni Research Foundation,
Center for High Throughput Computing (CHTC) at the University of Wisconsin{\textendash}Madison,
Open Science Grid (OSG),
Extreme Science and Engineering Discovery Environment (XSEDE),
Frontera computing project at the Texas Advanced Computing Center,
U.S. Department of Energy-National Energy Research Scientific Computing Center,
Particle astrophysics research computing center at the University of Maryland,
Institute for Cyber-Enabled Research at Michigan State University,
and Astroparticle physics computational facility at Marquette University;
Belgium {\textendash} Funds for Scientific Research (FRS-FNRS and FWO),
FWO Odysseus and Big Science programmes,
and Belgian Federal Science Policy Office (Belspo);
Germany {\textendash} Bundesministerium f{\"u}r Bildung und Forschung (BMBF),
Deutsche Forschungsgemeinschaft (DFG),
Helmholtz Alliance for Astroparticle Physics (HAP),
Initiative and Networking Fund of the Helmholtz Association,
Deutsches Elektronen Synchrotron (DESY),
and High Performance Computing cluster of the RWTH Aachen;
Sweden {\textendash} Swedish Research Council,
Swedish Polar Research Secretariat,
Swedish National Infrastructure for Computing (SNIC),
and Knut and Alice Wallenberg Foundation;
Australia {\textendash} Australian Research Council;
Canada {\textendash} Natural Sciences and Engineering Research Council of Canada,
Calcul Qu{\'e}bec, Compute Ontario, Canada Foundation for Innovation, WestGrid, and Compute Canada;
Denmark {\textendash} Villum Fonden and Carlsberg Foundation;
New Zealand {\textendash} Marsden Fund;
Japan {\textendash} Japan Society for Promotion of Science (JSPS)
and Institute for Global Prominent Research (IGPR) of Chiba University;
Korea {\textendash} National Research Foundation of Korea (NRF);
Switzerland {\textendash} Swiss National Science Foundation (SNSF);
United Kingdom {\textendash} Department of Physics, University of Oxford.

\end{document}